\documentclass[epj]{svjour}
\usepackage{graphics,amsmath,amssymb,multirow,color,pstricks,pst-plot}
\usepackage{xspace,graphicx,comment,hyperref,cite}

\newlength{\digitwidth} 

\newcommand{\pTovM}{\ensuremath{p_{\rm T}/M}\xspace}
\newcommand{\jpsi}{\ensuremath{{\rm J}/\psi}\xspace}
\newcommand{\psip}{\ensuremath{\psi{\rm (2S)}}\xspace}
\newcommand{\chic}{\ensuremath{\chi_{c}}\xspace}
\newcommand{\chicOne}{\ensuremath{\chi_{c1}}\xspace}
\newcommand{\chicTwo}{\ensuremath{\chi_{c2}}\xspace}
\newcommand{\upsOne}{\ensuremath{\Upsilon{\rm (1S)}}\xspace}

\newcommand{\oneSzero}{\ensuremath{{^1{\rm S}_0^{[8]}}}\xspace}
\newcommand{\threeSone}{\ensuremath{{^3{\rm S}_1^{[8]}}}\xspace}
\newcommand{\threePJ}{\ensuremath{{^3{\rm P}_J^{[8]}}}\xspace}

\newcommand{\pt}{\ensuremath{p_{\rm T}}\xspace}
\newcommand{\chictwooverchicone}{\ensuremath{\chi_{c2}/\chi_{c1}}\xspace}

\begin{document}

\title{From identical S- and P-wave $p_{\rm T}/M$ spectra\\
to maximally distinct polarizations:\\
probing NRQCD with $\chi$ states}

\titlerunning{Probing NRQCD with $\chi$ states}

\authorrunning{P. Faccioli et al.}

\author{Pietro Faccioli\inst{1} 
\and Carlos Louren\c{c}o\inst{2}
\and Mariana Ara\'ujo\inst{2}
\and Jo\~ao Seixas\inst{1}
\and Ilse Kr\"atschmer\inst{3}
\and Valentin Kn\"unz\inst{2}
}
\institute{LIP and IST, Lisbon, Portugal
\and 
CERN, Geneva, Switzerland
\and
HEPHY, Vienna, Austria
}

\date{Received: September 27, 2017 / Revised version: date}
\abstract{
A global analysis of ATLAS and CMS measurements reveals that, at mid-rapidity,
the directly-produced $\chi_{c1}$, $\chi_{c2}$ and J/$\psi$ mesons
have differential cross sections of seemingly identical shapes, 
when presented as a function of the mass-rescaled transverse momentum, \pTovM.
This identity of kinematic behaviours among S- and \mbox{P-wave} quarkonia
is certainly not a natural expectation of non-relativistic QCD (NRQCD), 
where each quarkonium state is supposed to reflect a specific family of 
elementary production processes, of significantly different \pt-differential cross sections.
Remarkably, accurate kinematic cancellations among the variegated NRQCD terms
(colour singlets and octets) of its factorization expansion
can lead to a surprisingly good description of the data.
This peculiar tuning of the NRQCD mixtures leads to a clear prediction regarding
the $\chi_{c1}$ and $\chi_{c2}$ polarizations, the only observables not yet measured:
they should be almost maximally different from one another, and from the J/$\psi$ polarization,
a striking exception in the global panorama of quarkonium production.
Measurements of the difference between the $\chi_{c1}$, $\chi_{c2}$ and J/$\psi$ polarizations, 
complementing the observed identity of momentum dependences, represent 
a decisive probe of NRQCD.
 % end of PACS codes
}

\maketitle

%Synopsis:
%Quarkonium measurements at the LHC show remarkably universal production and polarization patterns. 
%Our data-driven analysis shows that the measurements are surprisingly well described by NRQCD, 
%through several seemingly coincidental cancellations among its production processes.
%This study shows that accurate measurements of chi_c1 and chi_c2 polarizations will 
%either confirm the conceptual validity of NRQCD or challenge its naturalness and reveal simpler conceptual approaches.

%%%%%%%%%%%%%%%%%%%%%%%%%%%%%%%%%%%%
\section{Introduction}
%%%%%%%%%%%%%%%%%%%%%%%%%%%%%%%%%%%%

The mechanisms behind hadron production continue to challenge our understanding: 
analytical perturbative QCD calculations are insufficient to tackle all the aspects 
of the strong interactions driving the binding of quarks into observable particles.
%
%Studies of quarkonium production can provide crucial progress 
%towards solving this problem and non-relativistic QCD (NRQCD)~\cite{bib:NRQCD} 
%is generally considered as the best theory approach in this area of 
%QCD phenomenology~\cite{bib:Brambilla:2010cs}.
%
Studies of quarkonium production can provide crucial progress 
towards solving this problem~\cite{bib:Brambilla:2010cs}.
According to non-relativistic QCD (NRQCD)~\cite{bib:NRQCD},
one of the theory approaches in this area of QCD phenomenology,
%In this framework, 
S- and \mbox{P-wave} quarkonia are produced from the binding of quark-antiquark pairs 
created with a variety of quantum numbers, in color singlet or octet configurations.
These terms are characterized by significantly different kinematic dependences and polarizations, 
determined by the short-distance cross sections (SDCs),
presently calculated at next-to-leading order 
(NLO)~\cite{bib:BKNPB,bib:BKMPLA,bib:Chao:2012iv,bib:Gong:2012ug}.
They contribute with probabilities proportional to long distance matrix elements (LDMEs), 
extracted from fits to experimental data.
While conceptually appealing and successful in several respects, 
it has been confusing to see that different groups performing global fits 
to experimental data extract significantly different matrix elements,
despite using identical theory 
calculations, as a result of using different data fitting 
strategies~\cite{bib:BKNPB,bib:BKMPLA,bib:Chao:2012iv,bib:Gong:2012ug}.
These puzzles and a potential solution were discussed in Ref.~\cite{bib:FaccioliPLB736},
mostly devoted to the quarkonia least affected by feed-down contributions, the \psip\ and $\Upsilon$(3S) states.
A detailed data-driven analysis of the cross sections and polarizations of five S-wave and two P-wave states,
complemented by an original comparison with theory calculations, was presented in Ref.~\cite{bib:FaccioliPLB773}.
That analysis is extended in this paper to address two main questions:
how different and experimentally recognizable are the $\chi_c$ production mechanisms
with respect to those of the \jpsi\ and \psip\ mesons;
and how important will be, for the understanding of quarkonium production,
new or improved $\chi_c$ measurements.

%%%%%%%%%%%%%%%%%%%%%%%%%%%%%%%%%%%%
\section{Data-driven considerations}
%%%%%%%%%%%%%%%%%%%%%%%%%%%%%%%%%%%%

\begin{figure}[t]
\centering
\includegraphics[width=0.97\linewidth]{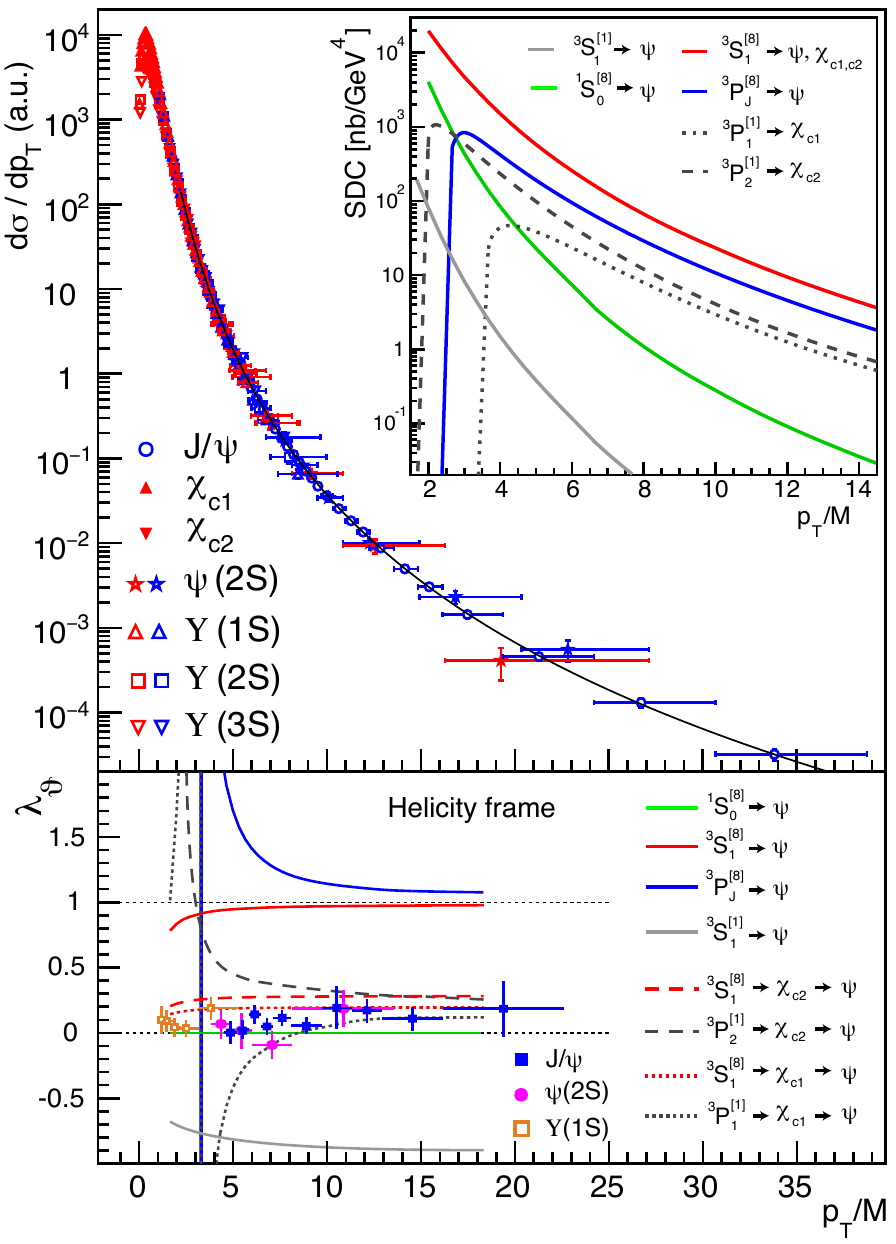}
\caption{Top: Mid-rapidity prompt quarkonium cross sections measured in pp collisions
at $\sqrt{s} = 7$\,TeV
by ATLAS (red markers)~\cite{bib:ATLASpsi2S, bib:ATLASYnS, bib:ATLASchic}
and CMS (blue markers)~\cite{bib:CMSjpsi, bib:CMSYnS}.
The normalizations were adjusted to the \jpsi\ points
to directly illustrate the universality of the \pTovM dependence.
%The curve represents a fit to all points of $\pTovM > 2$~\cite{bib:FaccioliPLB736},
The curve represents a fit to all points of $\pTovM > 2$~\cite{bib:FaccioliPLB773},
with a normalized $\chi^2$ of $215/193 = 1.11$.
The inset shows the NLO SDCs~\cite{bib:Chao:2012iv, bib:Shao:2014fca, bib:Shao:2015vga}.
The \threePJ and ${^3{\rm P}_{1,2}^{[1]}}$ SDCs are multiplied by $m_c^2$,
the mass of the charm quark squared;
they are negative and plotted with flipped signs.
Bottom: Polar anisotropy parameter $\lambda_\vartheta$, in the helicity frame,
measured by CMS in pp collisions at $\sqrt{s} = 7$\,TeV,
for prompt \jpsi, \psip and \upsOne dimuon decays~\cite{bib:CMSlambdaPsi2S,bib:CMSlambdaYnS}.
For improved visibility, values corresponding to two or three rapidity bins were averaged.
The curves represent the calculated 
$\lambda_\vartheta = 
(\mathcal{S}_{\rm T}-\mathcal{S}_{\rm L}) / 
(\mathcal{S}_{\rm T}+\mathcal{S}_{\rm L})$ values,
where $\mathcal{S}_{\rm T}$ ($\mathcal{S}_{\rm L}$) 
is the transverse (longitudinal) short distance cross section,
in the helicity frame (HX).}
%The curves represent the calculated $\lambda_\vartheta$ values.}
\label{fig:universal}
\end{figure}

As shown in the top panel of Fig.~\ref{fig:universal},
the ${^3{\rm S}_1}$ and ${^3{\rm P}_J}$ charmonium and bottomonium cross sections 
measured at the LHC at mid-rapidity show a remarkably uniform pattern as a function of \pTovM,
the ratio between the quarkonium transverse momentum and its mass
(Fig.~1 of Ref.~\cite{bib:FaccioliPLB773} shows the seven independent distributions).
Moreover, the corresponding measurements of the quarkonium decay distributions
indicate similar polarizations
% (all compatible with zero) 
for all S-wave states,
independently of their different \mbox{P-wave} feed-down contributions, 
as expressed by their polar anisotropy parameters, in the helicity frame, 
shown in the bottom panel.
%as expressed by the polar anisotropy parameters $\lambda_\vartheta$, 
%in the helicity frame,
%shown in the bottom panel (all compatible with zero).
%

This seemingly ``universal'' picture of quarkonium production is an unexpected result,
when compared to the wide variety of kinematic shapes of the differential cross sections (NLO SDCs)
contributing to the observable patterns within the NRQCD framework,
as shown in the inset of Fig.~\ref{fig:universal}.
The most surprising aspect is that the \chicOne\ and \chicTwo\ \mbox{P-wave} states have, 
at least at mid-rapidity, \pTovM distribution shapes indistinguishable from those of the \mbox{S-wave} states. 
According to the SDCs calculated at NLO, on the other hand, 
the singlet and octet \mbox{P-wave} terms of the NRQCD expansion, 
which contribute differently to \jpsi\ (\psip), \chicOne\ and \chicTwo\ production, 
have rather peculiar and differentiated kinematic behaviours,
with cross section terms becoming negative above characteristic \pTovM\ thresholds
and having unphysical polarization parameters ($|\lambda_\vartheta|>1$). 
In advance of any detailed numerical analysis, the qualitative comparison between data and theory 
illustrated in Fig.~\ref{fig:universal} indicates that the theory requires precise 
and seemingly unnatural cancellations between terms of the expansion, 
in order to reproduce observable cross sections and polarizations that are not only physical 
but also identical for states of different quantum numbers.

It should be noted that comparing the shapes of seven different quarkonium states,
including five \mbox{S-wave} states affected by very different fractions of \mbox{P-wave}
feed-down contributions, provides a stronger (more precise) statement regarding the overall equality
between \mbox{S-} and \mbox{P-wave} quarkonium production than one might initially expect,
given the uncertainties of the \chicOne\ and \chicTwo\ measurements on their own.
We will quantify this observation when presenting the results of the charmonium fit.

\begin{figure}[t]
\centering
\includegraphics[width=0.89\linewidth]{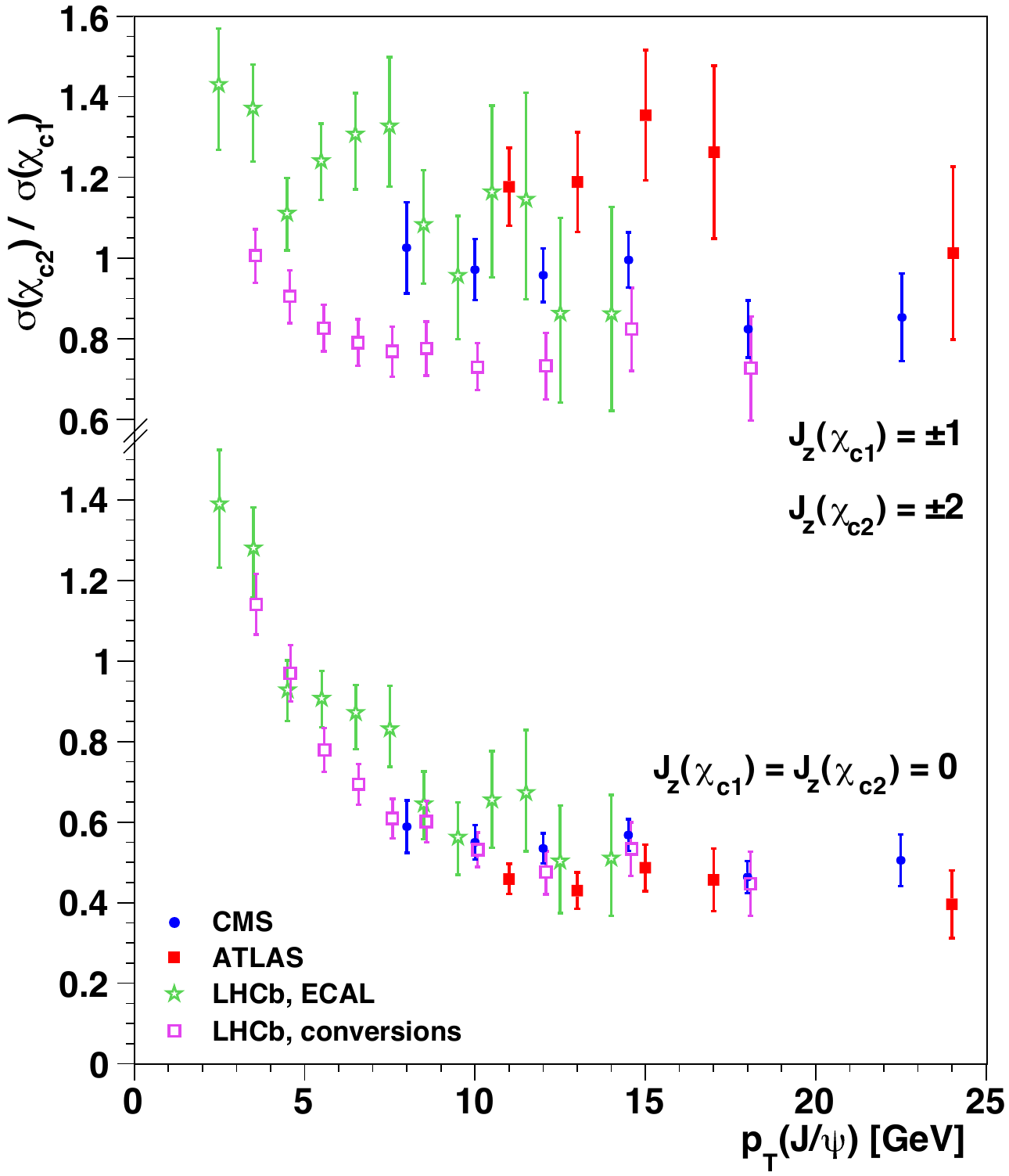}
\caption{The \chictwooverchicone\ ratio measured in pp collisions at 7~TeV
by ATLAS~\cite{bib:ATLASchic}, CMS~\cite{bib:CMSchic}
and LHCb~\cite{bib:LHCb_chicRatio,bib:LHCb_chicRatio_conversions}, 
with acceptance corrections calculated with two extreme polarization hypotheses:
spin alignments $J_z(\chi_{c1}) = \pm 1$, $J_z(\chi_{c2}) = \pm 2$ (top) 
and $J_z(\chi_{c1}) = J_z(\chi_{c2}) = 0$ (bottom).
The unpolarized hypothesis leads to intermediate values.}
\label{fig:chiRatioPol}
\end{figure}

The \chicOne\ and \chicTwo\ polarizations are
the main missing element in the current experimental landscape, but
two data-driven observations provide indirect indications.
First, the \psip, \jpsi and \upsOne polarizations are very similar (Fig.~\ref{fig:universal}-bottom), 
despite the diversity of $\chi$ feed-down fractions 
(0, $\sim$\,25\%~\cite{bib:ATLASchic,bib:FLSW_feedown} 
and $\sim$\,40\%~\cite{bib:LHCbChibFeedown}, respectively).
%suggesting that the sum of the \chicOne\ and \chicTwo\ 
%contributions has a negligible polarization.
%
Assuming that the \emph{directly-produced} \mbox{S-wave} mesons 
have very similar production mechanisms,
as indicated by the seemingly identical shapes of the \pTovM-differential cross sections
(Fig.~\ref{fig:universal}-top),
the \chicOne\ plus \chicTwo\ summed feed-down contributions 
cannot have a large impact in the observed \jpsi polarization. 
The second observation derives from comparing \chictwooverchicone\ cross-section ratios
measured in different experimental acceptances,
profiting from their strong sensitivity
to the polarization hypothesis used in the acceptance corrections.
Interestingly, as seen in Fig.~\ref{fig:chiRatioPol}, the $J_z = 0$ alignment hypothesis 
gives the best mutual agreement between the
\chictwooverchicone\ ratios reported by ATLAS and CMS, as well as
between the LHCb values obtained using photons detected in the 
calorimeter or with conversions to $e^+e^-$ pairs.
When both are polarized in the $J_z = 0$ limit, 
the $\chi_{c1}$ and $\chi_{c2}$ decays produce strongly polarized \jpsi mesons, with, 
respectively, $\lambda_\vartheta = +1$ and $\lambda_\vartheta = - 3/5$, 
leading to a weighted $\lambda_\vartheta \sim 0.3$ 
when the feed-down fractions and the cross-section ratio itself are taken into account. 
These observations suggest that the $\chi_{c1}$ and $\chi_{c2}$ 
polarizations might be a striking exception in the global 
panorama of high-energy quarkonium production, at least at mid rapidity.

%%%%%%%%%%%%%%%%%%%%%%%%%%%%%%%%%%%%
\section{Analysis method}
%%%%%%%%%%%%%%%%%%%%%%%%%%%%%%%%%%%%

To quantify our previous data-driven considerations and compare the results with theory, 
we perform a simultaneous fit of the mid-rapidity differential cross sections and polarizations,
including a detailed account of how the mother's 
momentum and polarization are transferred to the daughter
in the relevant feed-down decays:
$\psip \to \chi_{c1,2} \; \gamma$;
$\psip \to \jpsi \; X$;
$\chi_{c1,2} \to \jpsi \; \gamma$.
The analysis is restricted to the charmonium family,
given the lack of experimental information on bottomonium feed-down fractions.
The rule for the momentum propagation from mother to daughter is, approximately,
$p_{\rm T}/m = P_{\rm T}/M$, where $M$ ($m$) and $P_{\rm T}$ ($p_{\rm T}$) are, respectively,
the mass and laboratory transverse momentum of the mother (daughter) particle~\cite{bib:FaccioliPLB773}.
The polarization transfer rules were calculated in the electric dipole approximation and
precisely account for the observable dilepton distribution with no need of higher-order terms~\cite{bib:chiPol}.
The fit is exclusively based on empirical parametrizations. 
Perturbative calculations of the production kinematics are not used as ingredients anywhere in our analysis, 
the outcome of the fit being exclusively determined by the measurements and, therefore, only affected by
statistical and systematic experimental uncertainties.

Inspired by the pattern of slightly transverse polarizations seen in Fig.~\ref{fig:universal},
we parametrize the directly-produced \jpsi and \psip\ cross section shapes
as a superposition of unpolarized ($\lambda_\vartheta=0$)
and transversely polarized ($\lambda_\vartheta=+1$) processes,
$\lambda_\vartheta$ being the polar anisotropy parameter of the dilepton decay
in the helicity frame~\cite{bib:EPJC69}:
$\sigma_{\rm dir} \propto [ (1-f_{\rm p}) \, g_{\rm u} + f_{\rm p} \, g_{\rm p} ]$,
where $f_{\rm p}$,
identical for the two charmonia,
is the fractional contribution of the polarized process considered 
at an arbitrary reference point $(p_{\rm T}/M)^{*}$.
The shape functions $g_{\rm u}(\pTovM)$ and $g_{\rm p}(\pTovM)$ describe the \pTovM\ dependences of, 
respectively, the unpolarized and polarized yields. 
Both are normalized to unity at the chosen $(p_{\rm T}/M)^{*}$:
$g(\pTovM) = h(\pTovM) / h((p_{\rm T}/M)^{*})$, with
%
%%\begin{linenomath}
\begin{equation}
\label{eq:powerLaw}
h(\pTovM) =
\frac{p_{\rm T}}{M} \cdot \bigg( 1+\frac{1}{\beta-2} \cdot \frac{(p_{\rm T}/M)^2}{\gamma} \bigg)^{-\beta}\, .
\end{equation}
%\end{linenomath}
%
The parameter $\gamma$ (having the meaning of the average \pTovM\ squared)
defines the function in the low-\pt\ turn-on region and is 
only mildly sensitive to the data we are considering here; 
hence, in the fit we consider $\gamma$ as a common free parameter.
The $\beta$ power-law exponent, instead, characterizes the high-\pt shape:
$h \propto (p_{\rm T}/M)^{1-2\beta}$ for $\pTovM \gg \sqrt{\gamma (\beta-2)}$.
Therefore, we distinguish the unpolarized and polarized cross sections with two different powers, 
$\beta_{\rm u}$ and $\beta_{\rm p}$, respectively,
identical for the \jpsi and \psip.
The relative contributions and shapes of the $g_{\rm u}$ and $g_{\rm p}$ functions
are constrained by the polarization data. 
In fact, the polarized yield fraction, equal to $f_{\rm p}$ at $(p_{\rm T}/M)^{*}$, 
can be expressed as a function of \pTovM as $3 \lambda_\vartheta(\pTovM)  / [4 - \lambda_\vartheta(\pTovM) ]$.

For the \chicOne\ and \chicTwo\ direct cross sections
we use the same general \pTovM\ shape parametrization, 
but without discriminating between polarized and unpolarized contributions,
which, in the absence of $\chi$ polarization data, would not be individually constrained by the fit.
In short, we consider four contributions to direct quarkonium production,
the unpolarized and polarized $\psi$ terms plus the \chicOne\ and \chicTwo\ cross sections,
altogether characterized by one $\gamma$ and four $\beta$ parameters,
$\beta_{\rm u}$, $\beta_{\rm p}$, $\beta(\chi_1)$ and $\beta(\chi_2)$.
Their theoretical counterparts are, respectively, \oneSzero, $\threeSone+\threePJ$,
$\threeSone+{^3{\rm P}_{1}^{[1]}}$ and $\threeSone+{^3{\rm P}_{2}^{[1]}}$
(where each term indicates the SDC function times the LDME constant),
the four leading cross section components foreseen by NRQCD hierarchies
for ${^3{\rm S}_1}$ and ${^3{\rm P}_J}$ quarkonium production.
However, this parallelism is only a guidance in the parametrization of the fit, not a theoretical input.
As discussed in more detail in Ref.~\cite{bib:FaccioliPLB773}, 
our approach is very different with respect to fits using the calculated SDC shapes,
where the fit results are mostly determined by the \pt-differential cross sections;
the less precise polarization data are not included in the fits or have a negligible effect. 
In our fit, the polarization data, versus \pTovM, 
have the exclusive role of constraining both the relative normalizations
and the differences in momentum dependence of the 
polarized and unpolarized 
%%%different process 
contributions.
The precision of these data-driven results will evolve as new measurements become available,
remaining insensitive to specific theoretical calculations and uncertainties.

Without $\chi$ polarization measurements,
the \chicOne\ and \chicTwo\ cross sections cannot help, today,
discriminating the $g_u$ and $g_p$ contributions to \jpsi\ production and, therefore, 
relate the parametrized direct-\jpsi\ polarization to the measured prompt one.
However, the two data-driven observations mentioned above
allow us to implement such a relation by adopting
an approximate constraint on the total $\chi_{c}$ polarization contribution
to \jpsi production.
%$0 < \lambda_\vartheta^{\jpsi \leftarrow \chi_c} < 0.3$.
%
Given that, on average, 
$\lambda_\vartheta^{\jpsi} \gtrsim \lambda_\vartheta^{\psip}$, 
we can infer that $\lambda_\vartheta^{\jpsi \leftarrow \chi_c}$ should be positive,
under the assumption that the direct \jpsi\ and \psip\ polarizations are equal.
On the other hand, the extreme hypothesis, discussed above, according to which
both \chicOne\ and \chicTwo\ are polarized in the $J_z = 0$ limit 
leads to $\lambda_\vartheta^{\jpsi \leftarrow \chi_c} \simeq 0.3$
(which, when weighted by the 25\% feed-down fraction of \jpsi from $\chi_c$, 
is comparable to the average difference 
$\lambda_\vartheta^{\jpsi} - \lambda_\vartheta^{\psip} \simeq 0.05$).
We can thus be confident that $\lambda_\vartheta^{\jpsi \leftarrow \chi_c}$
is positive and not larger than 0.3.
The results of the fit, and ensuing considerations,
are insensitive to variations of $\lambda_\vartheta^{\jpsi \leftarrow \chi_c}$ 
(assumed to be \pTovM-independent) within this range.

The ATLAS and CMS integrated-luminosity uncertainties are (independently)
varied as nuisance parameters, following Gaussian functions centred at unity 
and of widths equal to the relative uncertainties of the published luminosities. 
These two nuisance parameters multiply all the data points (cross sections) 
of the respective experiment, thereby correlating the several datasets within 
each experiment. 
Moreover, the experiments measured products of  cross sections times 
branching ratios, so that the uncertainties of the branching ratios have also 
been treated as nuisance parameters (with central values and uncertainties taken from Ref.~\cite{bib:PDG}),
multiplying all relevant data points and
representing correlations between ATLAS and CMS.

Another source of correlation between all the points being fitted
is the dependence of the detection acceptances on the polarization.
For each set of parameter values considered in the fit scan, the expected values 
of the polarizations and cross sections are calculated, for all states, as functions 
of \pt, using the shape-parametrization functions described above.
The expected $\lambda_\vartheta$ values can be immediately compared to the 
measured ones, for the determination of the corresponding $\chi^2$ terms, 
while for the calculation of the cross-section $\chi^2$ terms we first scale the
measured cross sections by acceptance-correction factors calculated for the
$\lambda_\vartheta$ value under consideration. 
These correction factors are computed, for each data point, using the tables 
published by the experiments (for exactly this purpose) for the cross sections of 
particles produced with fully transverse or fully longitudinal polarization.

The fit has 100 experimental constraints and 20 parameters: 5 shape parameters, 
4 normalizations and the fraction $f_p$, plus 2 luminosity and 8 branching-ratio 
nuisance parameters.

%%%%%%%%%%%%%%%%%%%%%%%%%%%%%%%%%%%%
\section{Analysis results}
%%%%%%%%%%%%%%%%%%%%%%%%%%%%%%%%%%%%

\begin{figure}[t]
\centering
\includegraphics[width=0.89\linewidth]{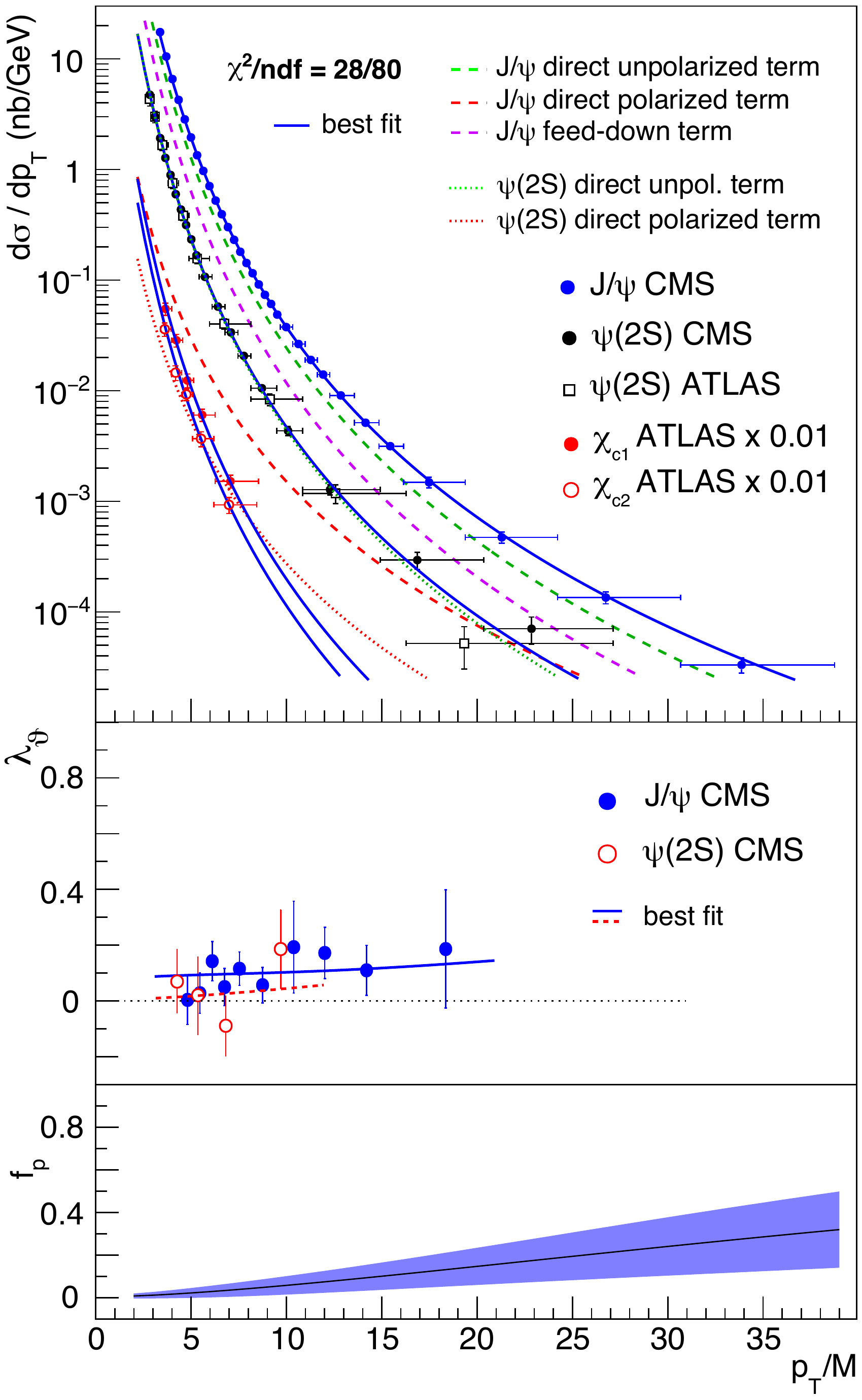}
\caption{Comparison between the data and the fitted curves,
for the \jpsi, \psip, \chicOne and \chicTwo cross sections (top) and
for the \jpsi and \psip polarizations (middle).
The bottom panel shows the resulting $\psi$ polarized fraction.}
\label{fig:fittedData}
\end{figure}

As shown in Fig.~\ref{fig:fittedData}, the charmonium cross sections and polarizations are
%very well 
described by the fit just presented, with a $\chi^2$ per degree of freedom of 28/80.
%, with the following parameters:
%$\gamma = 0.73 \pm 0.19$,
%$\beta_{\rm u}(\psi) = 3.42 \pm 0.05$,
%$\beta_{\rm p}(\psi) = 2.67 \pm 0.29$,
%$\beta(\chi_1) = 3.46 \pm 0.08$ and
%$\beta(\chi_2) = 3.49 \pm 0.10$.
%
\begin{figure}[t]
\centering
\includegraphics[width=0.9\linewidth]{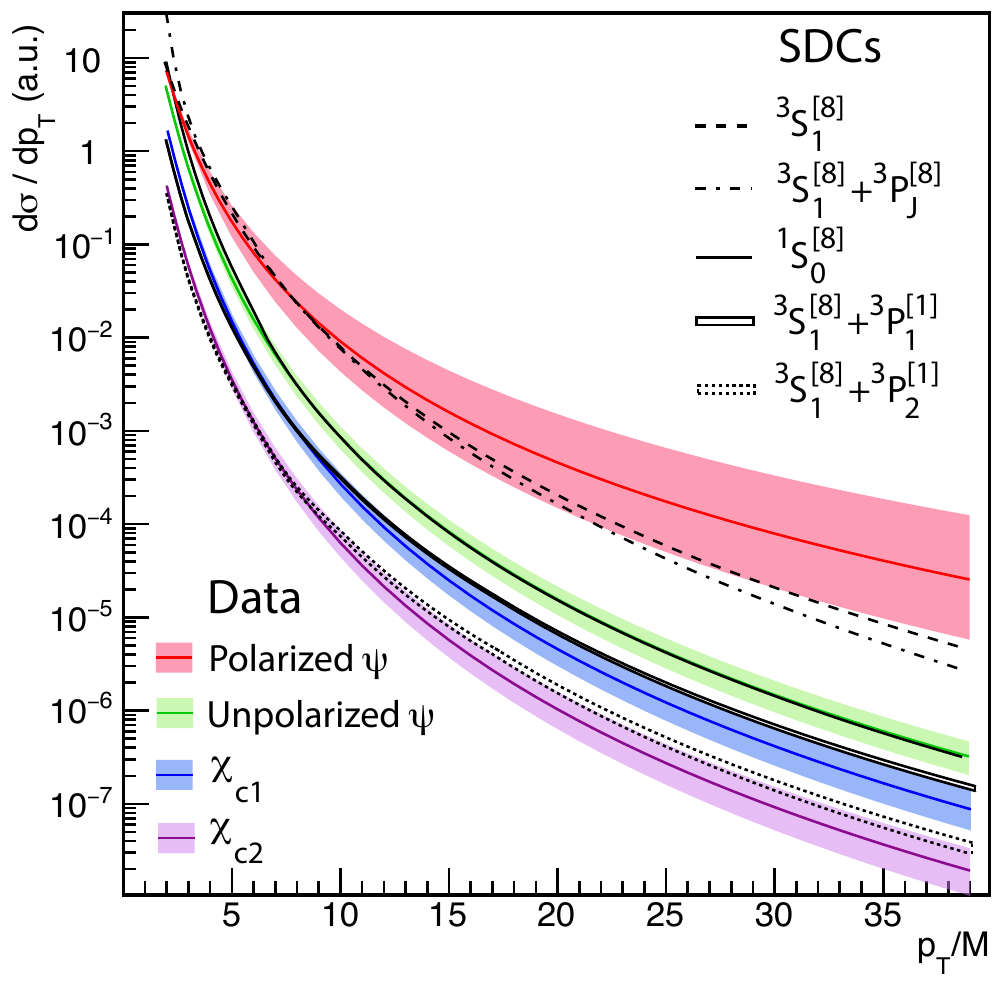}
\caption{Direct production cross sections resulting from the fit of the data,
with 68.3\% confidence level uncertainty bands reflecting correlated variations in the fit parameters.
The normalizations of the four bands are chosen for visibility reasons.
Suitable SDC combinations are also shown, 
normalized to the respective bands at $\pTovM = 8$.
The widths of the \chicOne and \chicTwo SDC bands 
reflect the 1.3\% uncertainty of $K_{\chi}$ (see text).}
\label{fig:datavsSDCs}
\end{figure}
Figure~\ref{fig:datavsSDCs} shows the fitted cross section terms
as bands of widths reflecting the experimental uncertainties.

A very interesting and non-trivial indication of this purely data-driven fit is that
the \chicOne and \chicTwo \pTovM\ distributions are very similar to the unpolarized term dominating $\psi$ production,
as quantified by the compatibility of the $\beta$ parameters: 
$\beta_{\rm u} = 3.42 \pm 0.05$, $\beta(\chi_1) = 3.46 \pm 0.08$ and $\beta(\chi_2) = 3.49 \pm 0.10$.
%all around 3.45.
%
This very clear experimental observation is predominantly the result of the 
perfect compatibility of the (high precision) \jpsi\ and \psip\ \pTovM\ shapes,
even in \pTovM\ ranges beyond those covered by the existing $\chi$ data,
reflecting the fact that the prompt \psip mesons are fully directly produced 
while $\simeq 25\%$~\cite{bib:ATLASchic} of the \jpsi\ yield comes from \chic\ decays.
In fact, the \chic\ cross sections are measured at relatively low \pt\ and with comparatively poor precision.
To verify this conclusion,
we repeated the fit keeping only one experimental point for each of the two \chic\ cross sections
(chosen in the middle of the measured range), 
so that these measurements constrain the feed-down fractions at that point but not the \pTovM\ shapes. 
As expected, the fit results for the \chic\ cross section shapes do not change significantly, 
with shape parameters remaining the same within the one-sigma range.

The experimental bands for the four observable cross sections are compared with the corresponding 
NRQCD terms, the dashed/dotted lines corresponding to (combinations of) SDCs calculated at NLO.
We emphasize that the two terms of comparison are completely independent,
the first being the result of a model-independent fit of experimental data
and the second a pure theoretical calculation.
The unpolarized and polarized $\psi$ bands are compared with, respectively,
the \oneSzero\ and \threeSone\ SDC shapes, calculated at NLO
and also including fragmentation corrections representing a partial account of 
next-to-next-to-leading order processes~\cite{bib:BodwinCorrections,bib:Bodwin:2015iua}.
As illustrated by the dot-dashed line, corresponding to
$(1/{m_c^2}) \;
\langle {\cal O}^{\jpsi} ({^3{\rm P}_0^{[8]}}) \rangle = 0.1 \cdot
\langle {\cal O}^{\jpsi} ({^3{\rm S}_1^{[8]}}) \rangle $,
adding a \threePJ\ term leads to steeper shapes,
departing from the polarized $\psi$ band more than the \threeSone\ term alone.
The \oneSzero SDC shape is in remarkable agreement with the experimental ``unpolarized'' band.
Moreover, adding the negative ${^3{\rm P}_{1,2}^{[1]}}$ SDCs to the \threeSone term
results in shapes approximating the \oneSzero term,
reproducing relatively well the observed similarity between the unpolarized-$\psi$ and
$\chi_{c1,2}$ patterns. 

Before discussing in more detail this data-theory comparison,
we will now describe the derivation of the NRQCD curves for
the $\chi_{c}$ distributions and corresponding polarization predictions.
In NRQCD the $\chi_{c1,2}$ polarizations and cross sections are functions of one common parameter, 
equal for all $\chi_c$ states,
\begin{equation}
\label{eq:Kchi}
K_{\chi} = ({1}/{m_c^2}) \, 
\langle {\cal O}^{\chi_{c0}} ({^3{\rm P}_0^{[1]}}) \rangle / 
\langle {\cal O}^{\chi_{c0}} ({^3{\rm S}_1^{[8]}}) \rangle \; ,
\end{equation}
with $\langle {\cal O} \rangle$
denoting the LDME.
The $\chi_c$ production cross sections $\sigma_J$
and the spin-density matrix elements $\sigma_J^{ij}$
have the general form
\begin{equation}
\label{eq:chiXsectionsNRQCD}
\sigma_J^{(ij)} \propto (2J+1) \left[ \mathcal{S}^{(ij)}({^3{\rm S}_1^{[8]}}) +  K_{\chi} \, m_c^2 \,\mathcal{S}^{(ij)}({^3{\rm P}_J^{[1]}})
\right] \;,
\end{equation}
where $\mathcal{S}^{(ij)}$ denotes the SDC or its spin projection.
The $\lambda_\vartheta$  are calculated as
$\lambda_\vartheta^{\chi_1} = ( \sigma_1^{00} - \sigma_1^{11} ) / ( \sigma_1^{00} + 3 \sigma_1^{11} )$
and 
$\lambda_\vartheta^{\chi_2} = 
( -3 \sigma_2^{00} - 3 \sigma_2^{11} + 6 \sigma_2^{22} ) / ( 5 \sigma_2^{00} + 9 \sigma_2^{11} + 6 \sigma_2^{22} )$,
where the $\sigma_J^{ij}$ depend on $K_{\chi}$ through 
Eq.~\ref{eq:chiXsectionsNRQCD}.
The $\chi_{c1,2}$ $\lambda_\vartheta$ parameters refer to the corresponding \jpsi\ dilepton decay distributions,
which are the ones directly measured and fully reflect the $\chi$ polarization state,
while being insensitive to the uncertain contributions of higher-order photon multipoles~\cite{bib:chiPol}.
%Moreover, the dilepton decay parameters are identical to the $\chi \to \jpsi\ \gamma$ ones, 

\begin{figure}[t]
\centering
\includegraphics[width=\linewidth,height=0.48\textheight]{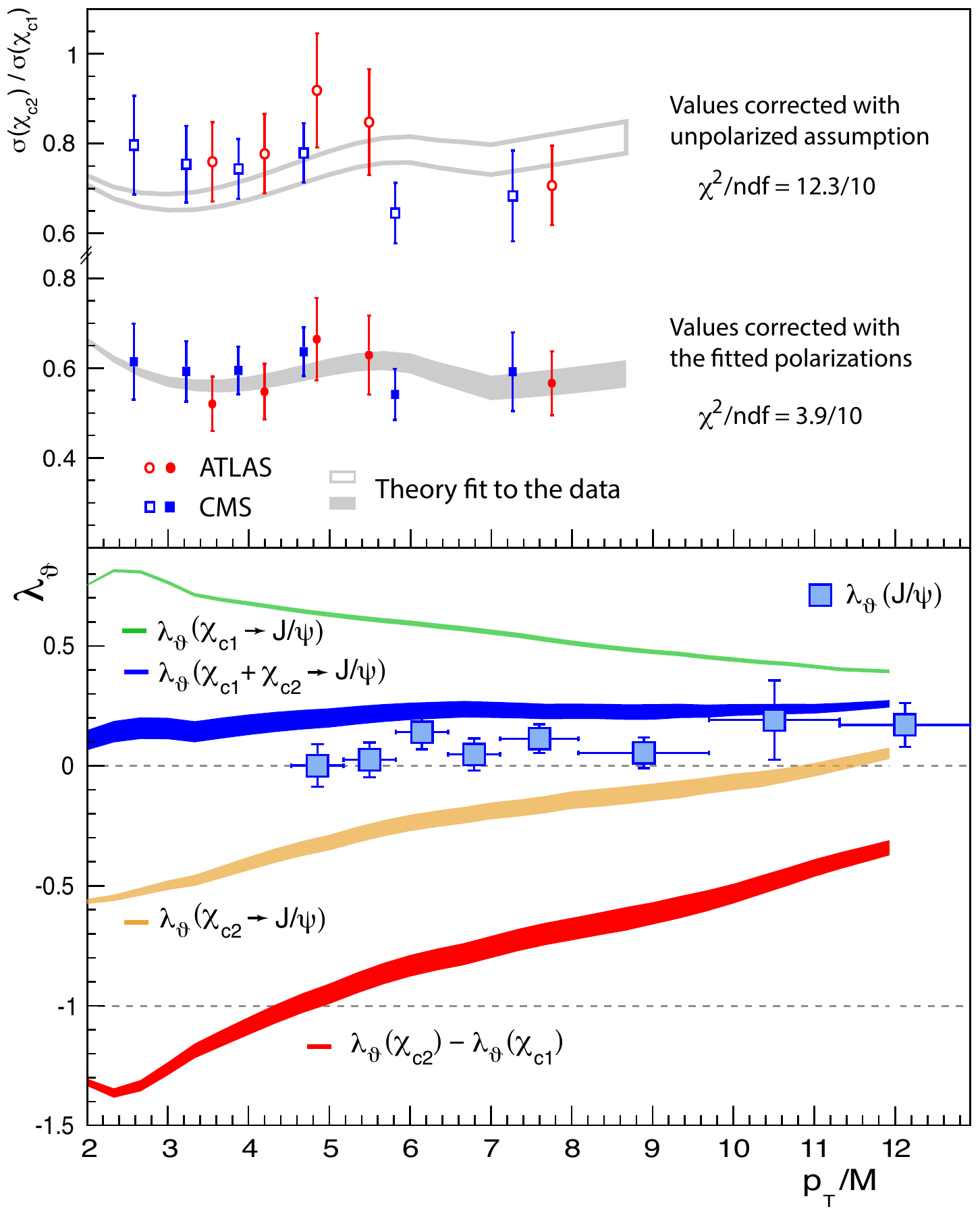}
\caption{Top: \chictwooverchicone\ ratio measured 
by ATLAS~\cite{bib:ATLASchic} and CMS~\cite{bib:CMSchic}, 
before (open markers) and after (filled markers) 
accounting for the dependence of the detection acceptances on the
(simultaneously calculated) \chicOne\ and \chicTwo\ polarizations.
The grey bands reflect the theory fits to the data, with
widths reflecting the $K_{\chi}$ uncertainty.
Bottom: $\chi_{c1,2}$ polarizations
calculated adding the \threeSone and ${^3{\rm P}_{1,2}^{[1]}}$ polarizations
with weights reflecting the \chictwooverchicone\ data.
The bands reflect correlated parameter variations.}
\label{fig:chiRatioFitNRQCD}
\end{figure}

We determine $K_{\chi}$ from the measured \chictwooverchicone\ ratio,
taking into account that the published values strongly depend on 
the \chicOne\ and \chicTwo\ polarizations assumed for the corrections
of the detector's acceptance. 
For each $K_{\chi}$ considered, 
we calculate the \chicOne\ and \chicTwo\ polarizations using NLO SDCs
and correct the published ratio by the corresponding acceptance ratio. 
The fit $\chi^2$ is then calculated comparing the corrected measurement 
(with statistical and systematic uncertainties, but no ``polarization uncertainties") 
with the prediction for that $K_{\chi}$ value.

Figure~\ref{fig:chiRatioFitNRQCD}-top shows how the \chictwooverchicone\ ratio changes, 
and the theoretical fit improves, 
when we use the NRQCD polarization conjecture instead of the unpolarized scenario 
that the experiments assume to report the measurements.
The result of our fit is $K_{\chi} = 4.60 \pm 0.06$,
much more precise than the value $3.7^{+1.0}_{-0.7}$,
derived~\cite{bib:Shao:2014fca}
using the unpolarized ratios and including the entire spectrum of polarization hypotheses 
in the experimental uncertainty.
The corresponding polarization predictions are shown in 
Fig.~\ref{fig:chiRatioFitNRQCD}-bottom.
Interestingly, as \pTovM\ decreases, $\lambda_\vartheta$ tends to the extreme physical values
$+1$ (\chicOne) and $-3/5$ (\chicTwo),
in agreement with the alignment scenario
suggested by the measured \chictwooverchicone\ cross-section ratios (Fig.~\ref{fig:chiRatioPol}):
these limit values correspond to two very different decay distribution shapes,
but to the same pure $J_z = 0$ angular momentum configuration of the $\chi_c$.
The \jpsi\ $\lambda_\vartheta$ from the weighted \chicOne and \chicTwo feed-downs
(blue band) is close to the values measured by CMS (squares) for the prompt sample, 
implying that the direct and feed-down terms have similar polarizations.

It is quite remarkable to observe that the difference 
$\Delta\lambda_\vartheta \equiv \lambda_\vartheta(\chi_{c2}) - \lambda_\vartheta(\chi_{c1})$
is predicted with a rather high precision and, furthermore,
reaches extreme values (around $-1$).
In particular, in the $\pt \approx 20$~GeV region, 
where experimental measurements will be provided in the near future, 
the prediction is $\Delta\lambda_\vartheta = -0.80 \pm 0.05$,
implying a strong deviation from the mild polarizations shown in Fig.~\ref{fig:universal}-bottom.
Comparing the discriminating power of 
this result to the corresponding predictions of Ref.~\cite{bib:Shao:2014fca} (Fig.~4),
$\lambda_\vartheta(\chi_{c1}) = 0.25^{+0.08}_{-0.05}$ and 
$\lambda_\vartheta(\chi_{c2}) = 0.10^{+0.15}_{-0.20}$,
one can see the crucial importance of a proper treatment of the uncertainties and
correlations affecting the experimental data.
It is also relevant to note that, 
thanks to the cancellation of most experimental systematic uncertainties,
the \emph{difference}
$\lambda_\vartheta(\chi_{c2}) - \lambda_\vartheta(\chi_{c1})$
can be measured with maximal significance and accuracy.

\begin{figure}[t]
\centering
\includegraphics[width=0.9\linewidth]{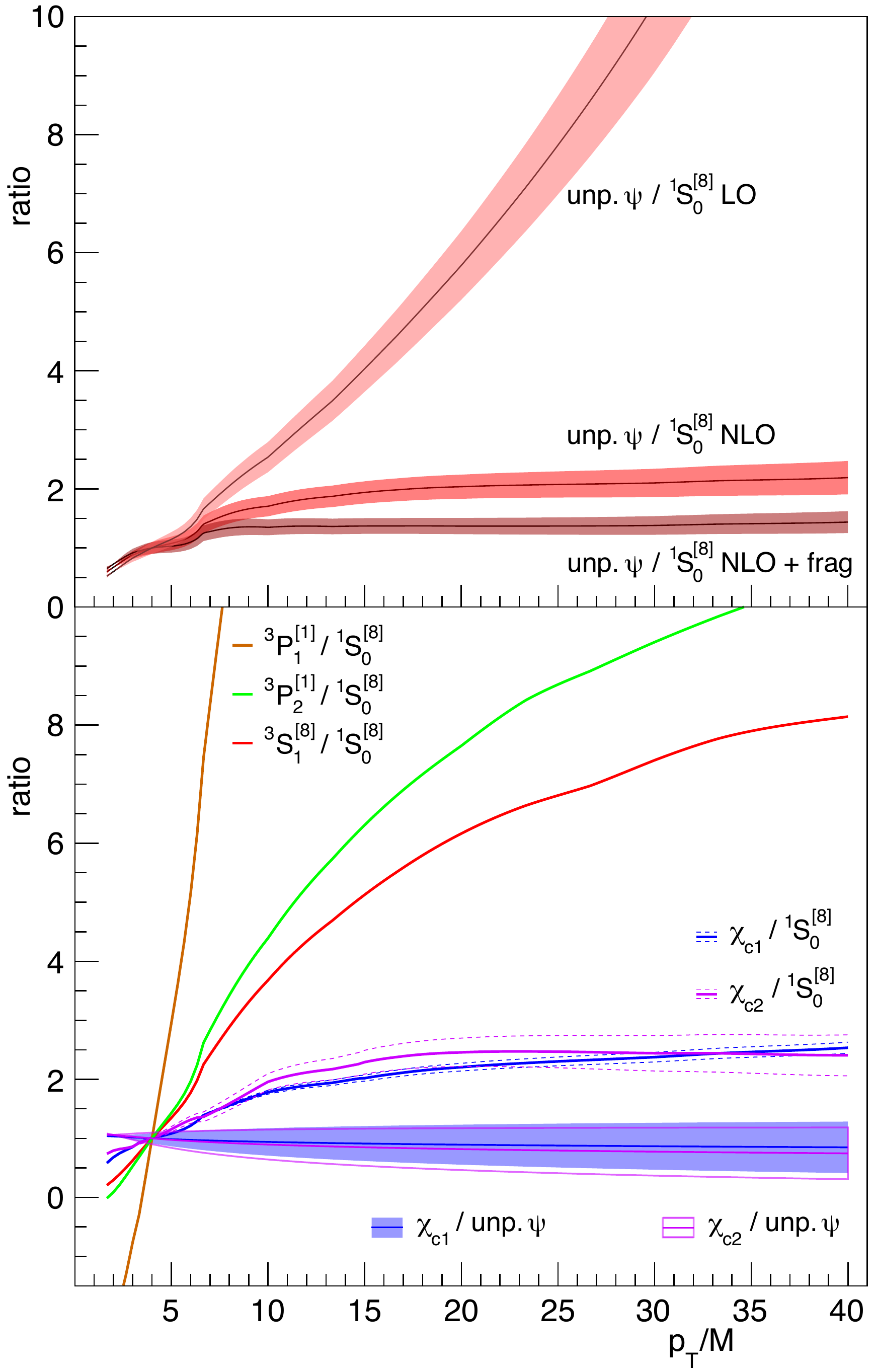}
\caption{Ratios of direct-production charmonium cross section shapes 
for different combinations of the measured and/or calculated terms 
already presented in Fig.~\ref{fig:datavsSDCs}. 
For visibility reasons, all ratios are normalized to unity at $\pTovM = 4$.}
\label{fig:datavsSDCs_ratios}
\end{figure}

We will now discuss in more detail the theory-data comparison. 
To discern shape differences more easily than in the logarithmic-scale plots of Fig.~\ref{fig:datavsSDCs},
we present in Fig.~\ref{fig:datavsSDCs_ratios} some of the results in the form of ratios, in a linear scale.
In the top panel, we can see that the ratio between the experimental ``unpolarized'' band 
and the state-of-the-art \oneSzero SDC curve (``NLO\,+\,frag." band)
only deviates from a perfectly flat function in the low-$\pTovM$ region ($\pTovM < 7$). 
This effect might represent a residual limitation of current finite-order perturbative calculations, 
as suggested by the observation that the ratio shows a more pronounced non-flatness 
when the SDC is calculated at NLO without fragmentation contributions (``NLO" band), 
and is not flat at all when we use the LO SDC as reference (``LO" band).
The differences between these three ratios provide a pedagogical illustration of the 
improvements made in the successive evolutions of the calculations.
The bottom panel shows that the ratios between the measured $\chi_{c1}$ (blue filled band) 
or $\chi_{c2}$ (pink open band) \pTovM-differential cross sections and the corresponding
unpolarized-$\psi$ cross section are practically identical to each other, and essentially flat,
offering an effective representation of the strong experimental observation mentioned above.
It is interesting to compare these two bands, exclusively determined by the measurements,
with the two corresponding (and completely independent) theory ratios,
here represented by the (blue and pink) solid lines,
calculated as the ratios between suitable combinations of the $\threeSone$ and ${^3{\rm P}_{1,2}^{[1]}}$ SDCs
(analogous to the $\chi_{c1}$ or $\chi_{c2}$)
and the \oneSzero SDC
(analogous to the unpolarized-$\psi$). 
The dashed curves surrounding the solid ones reflect the 1.3\% uncertainty on $K_{\chi}$,
already shown in Fig.~\ref{fig:datavsSDCs}.
Also this ratio deviates from a flat function in the lower part of the $\pTovM$ range,
but this deviation is a relatively small effect, as can be judged by comparing it with 
the corresponding rate of increase of the \emph{individual} components,
$\threeSone / \oneSzero$ (red curve) and ${^3{\rm P}_{1,2}^{[1]}}  / \oneSzero$ (brown and green curves). 
It is actually quite remarkable to see how effective is the mutual cancellation of the individual (steep) variations,
in the combinations pertinent to the $\chi_{c1}$ and $\chi_{c2}$ states.
Also taking into consideration that the \mbox{P-wave} SDCs seem to be affected 
by a slower convergence of the perturbative series than the S-wave SDCs~\cite{bib:Bodwin:2015iua}, 
the present level of agreement between the shapes of the $\chi_{c}$-to-\oneSzero predicted ratios 
and the corresponding $\chi_{c}$-to-unpolarized-$\psi$ measured bands can be considered very promising.
As a matter of fact, 
and despite the initial impression of unnecessary complexity expressed by Fig.~\ref{fig:universal},
we see that NRQCD provides predictions that are, already today, 
very close to reproducing the uniformity of the observed \pTovM trends, 
as well as the small measured \mbox{S-wave} polarizations.
This unexpected agreement is the result of a series of cancellations, which, 
given their fragile and unstable nature, must be tested with precise ingredients. 
Further improvements in the perturbative calculations, especially for the \mbox{P-wave} SDCs, 
are needed for more conclusive statements.

%%%%%%%%%%%%%%%%%%%%%%%%%%%%%%%%%%%%
\section{Summary}
%%%%%%%%%%%%%%%%%%%%%%%%%%%%%%%%%%%%

The $\chi_{c1}$ and $\chi_{c2}$ states have, both, \pTovM distributions 
well compatible with being identical to the one of the \jpsi mesons. 
This conclusion results from the study of the full set of charmonium data
and has a much higher significance than one would obtain if only considering the
$\chi_{c}$ cross section measurements, given their limited precision and \pt coverage
in comparison to the \jpsi\ and \psip\ measurements.
This is a very specific and non-trivial experimental observation, 
seemingly in contradiction, at least a priori, with the expectations of NRQCD,
given the significantly different shapes of the relevant SDCs. 
For example, the ratios between
the $\threeSone$ and ${^3{\rm P}_{1,2}^{[1]}}$ SDCs, dominant terms of the factorization expansion for $\chi$ production,
to the \oneSzero SDC, very well describing \jpsi\ and \psip\ production,
have an order-of-magnitude increase from low to high \pTovM.
Therefore, within NRQCD, one would a priori expect different \pTovM\ dependences for
the $\chi_{c1}$, $\chi_{c2}$ and $\psi$ states.
Remarkably, thanks to mutual cancellations of the steep SDC shapes differences,
NLO NRQCD calculations approximately reproduce the similarity betwen the $\chi_{c1}$, $\chi_{c2}$ and $\psi$ cross setions shapes,
giving a satisfactory description of charmonium production as
measured at mid-rapidity by the ATLAS and CMS experiments.

On the other hand, at the present state of the SDC calculations and
within the limits of the currently adopted $v^2$ expansion,
this agreement comes with a definite prediction
of strong and opposite \chicOne and \chicTwo polarizations.
It is worth restating this conclusion with different words.
The $\chi_{c1}$ and $\chi_{c2}$ differential cross sections have
\pTovM\ dependences compatible with being identical to the \jpsi distribution,
an experimental observation that NRQCD
(given the presently available NLO SDCs)
can only reproduce
in a very specific and non-trivial configuration,
leading to a remarkable prediction:
the $\chi_{c1}$ and $\chi_{c2}$ polarizations
are as different from each other as physically possible.

If confirmed experimentally, through an accurate measurement of the
variable $\lambda_\vartheta(\chi_{c2}) - \lambda_\vartheta(\chi_{c1})$,
the existence of strong $\chi_{c1}$ and $\chi_{c2}$ polarizations
(an exception among all quarkonia observed by high-\pt\ experiments)
would be a big step forward to confirm the existence of the
diversified and polarized processes that are at the heart of NRQCD.
If, instead, similar and weak \chicOne and \chicTwo polarizations will be measured,
it will be crucial to investigate if the predicted strong and opposite polarizations,
experimentally falsified, are caused by approximations and inaccuracies of the
presently available fixed-order perturbative calculations or
from problems in the conceptual foundations of the theory.
In that case, NRQCD would be facing a big challenge:
even if future improvements of the \mbox{P-wave} SDC calculations would eventually make
the $\chi_{c1}$ and $\chi_{c2}$ polarization predictions compatible with the measurements
(e.g., building upon the recent progress on fragmentation corrections~\cite{bib:Bodwin:2015iua})
one would still think that the homogeneity of the observed kinematic patterns deserves
a more natural theoretical explanation than a series of ``coincidences'' cancelling out the
variegated complexity of NRQCD.
In either case,
accurate measurements of the \chicOne\ and \chicTwo\ polarizations constitute a decisive test of NRQCD.

\begin{acknowledgement}

H.-S.\;Shao kindly provided the NLO SDC calculations.
The work of I.K.\ is supported by FWF, Austria,
through the grant  P\,28411-N36.

\end{acknowledgement}

%\bibliographystyle{cl_unsrt}
%\bibliography{GlobalFit2016}{}

\providecommand{\href}[2]{#2}
\begingroup
\raggedright

\endgroup

\end{document}